\documentclass[]{spie}  %>>> use for US letter paper
\usepackage{amsmath,amsfonts,amssymb}
\usepackage{comment}
\usepackage{graphicx}
\usepackage[colorlinks=true, allcolors=blue]{hyperref}

\title{Performance of near-infrared high-contrast imaging methods with JWST from commissioning}

\author[a]{Jens Kammerer}
\author[a]{Julien Girard}
\author[b]{Aarynn L. Carter}
\author[a]{Marshall D. Perrin}
\author[a]{Rachel Cooper}
\author[a]{Deepashri Thatte}
\author[c]{Thomas Vandal}
\author[d]{Jarron Leisenring}
\author[e,f]{Jason Wang}
\author[g,a]{William O. Balmer}
\author[a,g]{Anand Sivaramakrishnan}
\author[a]{Laurent Pueyo}
\author[h]{Kimberly Ward-Duong}
\author[a]{Ben Sunnquist}
\author[i]{J\'ea Adams Redai}
\affil[a]{Space Telescope Science Institute, 3700 San Martin Dr, Baltimore, MD 21218, USA}
\affil[b]{University of California Santa Cruz, Santa Cruz, CA 95064, USA}
\affil[c]{D\'epartement de Physique and Observatoire du Mont-M\'egantic, Universit\'e de Montr\'eal, C.P. 6128, Succ. Centre-ville, Montr\'eal, H3C 3J7, Qu\'ebec, Canada}
\affil[d]{Steward Observatory, University of Arizona, 933 N. Cherry Ave, Tucson, AZ 85721-0065, USA}
\affil[e]{Department of Astronomy, California Institute of Technology, Pasadena, CA 91125, USA}
\affil[f]{Center for Interdisciplinary Exploration and Research in Astrophysics (CIERA) and Department of Physics and Astronomy, Northwestern University, Evanston, IL 60208, USA}
\affil[g]{Department of Physics \& Astronomy, Johns Hopkins University, 3400 N. Charles Street, Baltimore, MD 21218, USA}
\affil[h]{Department of Astronomy, Smith College, Northampton, MA 01063, USA}
\affil[i]{Center for Astrophysics, Harvard \& Smithsonian, 60 Garden Street, Cambridge, MA 02138, USA}

\authorinfo{Further author information: send correspondence to jkammerer@stsci.edu}

\begin{document} 
\maketitle

\begin{abstract}
The \emph{James Webb Space Telescope} (\emph{JWST}) will revolutionize the field of high-contrast imaging and enable both the direct detection of Saturn-mass planets and the characterization of substellar companions in the mid-infrared. While \emph{JWST} will feature unprecedented sensitivity, angular resolution will be the key factor when competing with ground-based telescopes.
Here, we aim to characterize the performance of several extreme angular resolution imaging techniques available with \emph{JWST} in the 3--$5~\text{\textmu m}$ regime based on data taken during the instrument commissioning.
Firstly, we introduce custom tools to simulate, reduce, and analyze \emph{JWST} NIRCam and MIRI coronagraphy data and use these tools to extract companion detection limits from on-sky NIRCam round and bar mask coronagraphy observations. Secondly, we present on-sky \emph{JWST} NIRISS aperture masking interferometry (AMI) and kernel phase imaging (KPI) observations from which we extract companion detection limits using the publicly available \texttt{fouriever} tool.
Scaled to a total integration time of one hour and a target of the brightness of AB~Dor (W1 $\approx$ 4.4~mag, W2 $\approx$ 3.9~mag), we find that NIRISS AMI and KPI reach contrasts of $\sim7$--8~mag at $\sim70$~mas and $\sim9$~mag at $\sim200$~mas. Beyond $\sim250$~mas, NIRCam coronagraphy reaches deeper contrasts of $\sim13$~mag at $\sim500$~mas and $\sim15$~mag at $\sim2$~arcsec. While the bar mask performs $\sim1$~mag better than the round mask at small angular separations $\lesssim0.75$~arcsec, it is the other way around at large angular separations $\gtrsim1.5$~arcsec. Moreover, the round mask gives access to the full 360~deg field-of-view which is beneficial for the search of new companions.
We conclude that already during the instrument commissioning, \emph{JWST} high-contrast imaging in the L- and M-bands performs close to its predicted limits and is a factor of $\sim10$ times better at large separations than the best ground-based instruments operating at similar wavelengths despite its $>2$ times smaller collecting area.
\end{abstract}


\section{INTRODUCTION}
\label{sec:introduction}

While the number of detected exoplanets has recently passed the bar of 5'000,\footnote[1]{\url{https://exoplanetarchive.ipac.caltech.edu/}} observations of wide separation gas giants akin to Jupiter and Saturn and young planetary systems during formation remain rare [\citenum{bowler2016},\citenum{currie2022}]. Consequently, theories about the formation of these gas giants and our very own Solar System are still lacking observational evidence [\citenum{helled2021}]. This is because most of the known exoplanets have been detected using indirect techniques such as the transit or the radial velocity technique. These techniques are most sensitive to planets at small orbital separations and are more easily applicable to mature and quiescent stars [\citenum{fischer2014}]. On the contrary, the direct imaging technique is more sensitive to planets at large orbital separations and young ages when the planets are still glowing bright from their remaining formation heat [\citenum{bowler2016}]. However, resolving and detecting the light of a planet several ten to hundred-thousand times fainter than its host star is challenging and current 8--10~m-class adaptive optics-corrected ground-based telescopes struggle to reach the regime of true young Jupiter analogs [\citenum{nielsen2019},\citenum{vigan2021}].

The \emph{James Webb Space Telescope} (\emph{JWST}) is expected to breach this line and directly detect planets similar in mass to Jupiter and even Saturn for some of the near young moving group (YMG) members [\citenum{carter2021}]. However, one caveat of \emph{JWST} is its smaller mirror ($\sim6.5$~m) if compared to the largest ground-based telescopes which, in combination with its longer observing wavelength regime, results in a worse inner working angle (IWA) performance [\citenum{perrin2018}]. In addition, the NIRCam focal plane masks were designed with a very large IWA of $\sim6~\lambda/D$, where IWA is defined as the half width at 50\% throughput [\citenum{krist2010}]. This means that companion mass limits on the order of $1~\mathrm{M}_\text{Jup}$ can only be achieved at several tens of astronomical units. To push direct imaging with \emph{JWST} closer to the snow line, where gas giants are expected to be most common based on trends from radial velocity and direct imaging surveys [\citenum{fernandes2019},\citenum{fulton2021}], extreme angular resolution techniques will be a key to exploit the unprecedented sensitivity of the largest space telescope that mankind has ever built.

The high-contrast imaging modes available with \emph{JWST} are NIRCam and MIRI coronagraphy [\citenum{green2005},\citenum{boccaletti2015}], NIRISS aperture masking interferometry (AMI, [\citenum{artigau2014}]) and kernel phase imaging (KPI, Kammerer et al. 2022 in prep.) which is in principle applicable with all \emph{JWST} imagers and integral-field spectrographs. The working principle of coronagraphy is to use a focal plane mask that attenuates the on-axis light of the host star while letting through the off-axis light of faint companions. Onboard \emph{JWST}, this is achieved with Lyot and four-quadrant phase masks [\citenum{soummer2005},\citenum{rouan2000}]. Modern coronagraphs typically achieve contrasts of $>10$~mag beyond $\sim2$--$3~\lambda/D$. Instead, Fourier plane imaging techniques such as AMI and KPI aim to disentangle the light of faint companions from the light of the host star using interferometric techniques. AMI employs a non-redundant pupil plane mask to turn the telescope primary mirror into a sparse interferometric array [\citenum{tuthill2000}] while KPI considers interference between all parts of the full pupil [\citenum{martinache2010}]. These techniques typically achieve contrasts of $<10$~mag, but enable detections at small angular separations down to $\sim0.5~\lambda/D$, well inside the IWA of classical coronagraphs [\citenum{ireland2013},\citenum{ireland2016}]. In this work, we aim to compare the performance of several \emph{JWST} high-contrast imaging techniques in the 3--$5~\text{\textmu m}$ regime, where \emph{JWST} is expected to reach its best IWA. These techniques are
\begin{enumerate}
    \item NIRCam round and bar mask coronagraphy at $\sim3.4~\text{\textmu m}$,
    \item NIRISS aperture masking interferometry (AMI) at $\sim4.8~\text{\textmu m}$,
    \item NIRISS kernel phase imaging (KPI) at $\sim4.8~\text{\textmu m}$.
\end{enumerate}

The paper is structured as follows. Section~\ref{sec:observations} introduces the NIRCam coronagraphy (Section~\ref{sec:nircam_coronagraphy}) and the NIRISS AMI and KPI (Section~\ref{sec:niriss_ami_and_kpi}) commissioning data that is used for evaluating the performance of these modes in Section~\ref{sec:results_and_discussion}. A more detailed description of the AMI and the KPI data reduction process can be found in the NIRISS commissioning papers by Sivaramakrishnan et al. 2022 (in prep.) and Kammerer et al. 2022 (in prep.), respectively. Section~\ref{sec:methods} gives an overview of our customly developed and publicly available tools to simulate (Section~\ref{sec:simulating_nircam_coronagraphy_data_with_nirccos_and_pynrc}) and to reduce and analyze (Section~\ref{sec:reducing_and_analyzing_nircam_coronagraphy_data_with_spaceklip}) NIRCam coronagraphy data. A more detailed description of our reduction and analysis tools and how they can also be applied to MIRI coronagraphy data can be found in Carter et al. 2022 (in prep.). We summarize our findings and conclude in Section~\ref{sec:summary_and_conclusions}.

\section{OBSERVATIONS}
\label{sec:observations}

To evaluate the performance of extreme angular resolution imaging with \emph{JWST}, we here present commissioning observations using the NIRCam coronagraphy [\citenum{green2005}] and the NIRISS aperture masking interferometry (AMI, [\citenum{artigau2014}]) and kernel phase imaging (KPI, Kammerer et al. 2022 in prep.) modes. We aim to compare the IWA and the contrast performance of these different high-contrast imaging modes. AMI and KPI achieve the better resolution and we also compare them to NIRcam coronagraphy which operates at similar wavelength. MIRI observes at longer wavelengths and thus can reach lower mass planets at larger separations [\citenum{carter2021}] but it is beyond the scope of the work presented here. The \emph{JWST} data described here can be found at \url{http://dx.doi.org/10.17909/9cde-9b92}.

\subsection{NIRCam coronagraphy}
\label{sec:nircam_coronagraphy}

The NIRCam coronagraphy data presented here come from \emph{JWST} commissioning program 1441 (PI Julien Girard). This program aimed to verify the NIRCam coronagraphic mask suppression by observing HD~114174, a target with a known white dwarf companion at $\sim500$~mas separation, and several point-source reference stars to enable PSF subtraction methods. Data was taken with all five NIRCam coronagraphic masks but we here focus on the 335R round mask and the LWB bar mask data which are expected to deliver the smallest IWA between 3--$5~\text{\textmu m}$ (which is the same wavelength regime where NIRISS AMI and KPI are operating). This firstly enables a more direct comparison between the NIRCam and the NIRISS observing modes and secondly is the wavelength regime where \emph{JWST} is expected to be superior to ground-based telescopes due to the thermal background of the atmosphere. A more comprehensive analysis of the commissioning data from program 1441 can be found in Girard et al. 2022 [\citenum{girard2022}]. The observing log for the NIRCam coronagraphy data considered here can be found in Table~\ref{tab:nircam_coronagraphy}.

\begin{table}[ht]
\caption{Observing log of the NIRCam coronagraphy observations considered in this work. $T_\text{exp}$ denotes the effective integration time per exposure so that the total integration time can be obtained as $N_\text{exp}T_\text{exp}$.}
\label{tab:nircam_coronagraphy}
\begin{center}
\begin{tabular}{lllllll}
\rule[-1ex]{0pt}{3.5ex} Type & Observing date \& time & Target name & Filter & Mask & $N_\text{exp}$/$N_\text{int}$/$N_\text{group}$ & $T_\text{exp}$ [s] \\
\hline
\hline
\rule[-1ex]{0pt}{3.5ex} SCI & 2022-07-05T22:08:00 & HD~114174 & F335M & 335R & 2/63/10 & 3300.126 \\
\rule[-1ex]{0pt}{3.5ex} REF & 2022-07-05T20:26:56 & HD~111733 & F335M & 335R & 9/7/10 & 366.681 \\
\rule[-1ex]{0pt}{3.5ex} REF & 2022-07-06T04:52:25 & HD~115640 & F335M & 335R & 9/7/10 & 366.681 \\
\rule[-1ex]{0pt}{3.5ex} REF & 2022-07-06T08:06:41 & HD~116249 & F335M & 335R & 9/7/10 & 366.681 \\
\hline
\rule[-1ex]{0pt}{3.5ex} SCI & 2022-07-06T01:22:22 & HD~114174 & F335M & LWB & 2/80/10 & 1710.464 \\
\rule[-1ex]{0pt}{3.5ex} REF & 2022-07-05T18:48:57 & HD~111733 & F335M & LWB & 5/16/10 & 342.093 \\
\rule[-1ex]{0pt}{3.5ex} REF & 2022-07-06T06:19:39 & HD~115640 & F335M & LWB & 5/16/10 & 342.093 \\
\hline
\end{tabular}
\end{center}
\end{table}

The coronagraphy ramp (uncal) data were first processed with the \texttt{jwst}\footnote[2]{\url{https://github.com/spacetelescope/jwst}} stage 1 (detector level calibrations) and stage 2 (photometric calibrations) pipelines. Then, a PSF reference library was constructed from the point-source reference star observations and the Karhunen-Lo\`eve image projection (KLIP, [\citenum{soummer2012}]) method with one annulus and one subsection was used to compute and subtract an optimized reference PSF from the science images using \texttt{spaceKLIP} (see Section~\ref{sec:reducing_and_analyzing_nircam_coronagraphy_data_with_spaceklip}). In the remainder of this work, we always subtract 100 Karhunen-Lo\`eve (KL) modes from the science images. When computing the KLIP basis, we consider every integration as a separate reference PSF so that we end up with $>100$ reference PSFs in all cases. We note that the \texttt{jwst} stage 3 coronagraphy pipeline can also be used to perform PSF subtraction using the KLIP method, but we opted for using \texttt{spaceKLIP} to have more control over the PSF subtraction parameters and to get access to the \texttt{pyKLIP} routines for determining detection limits and extracting companion properties [\citenum{wang2015}].

\subsection{NIRISS AMI \& KPI}
\label{sec:niriss_ami_and_kpi}

\begin{figure}
\centering
\includegraphics[trim={2.00cm 4.75cm 2.00cm 9.25cm},clip,width=0.75\textwidth]{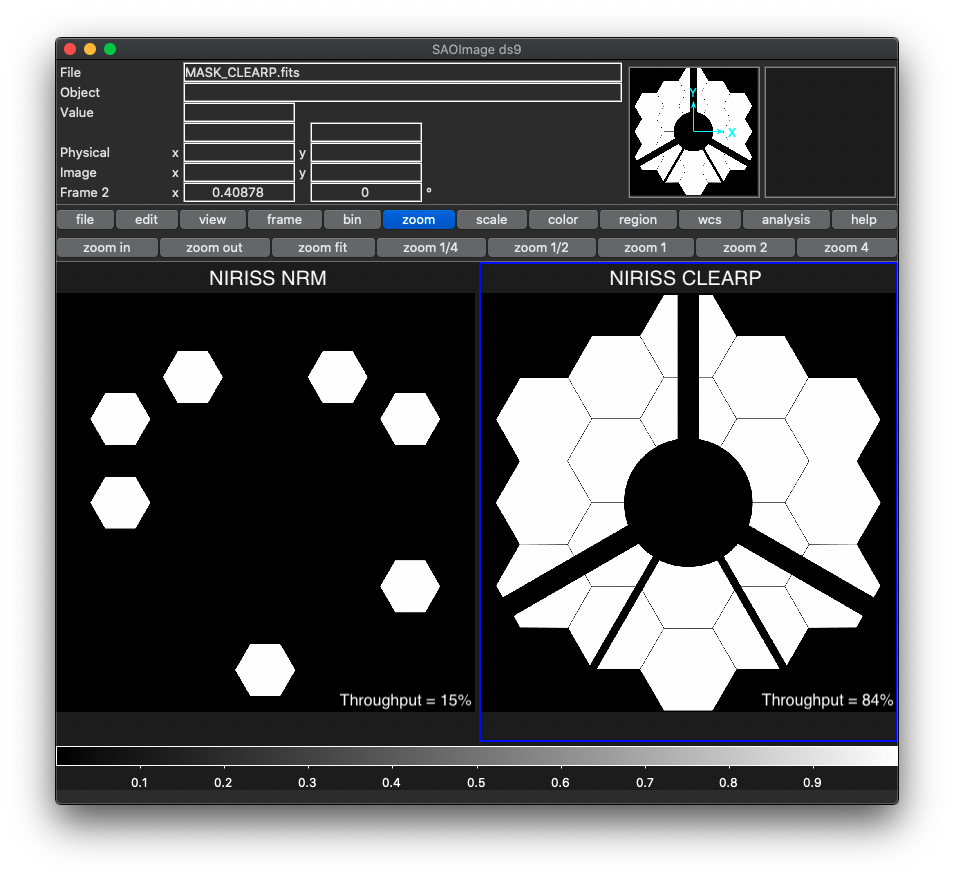}
\caption{NIRISS non-redundant mask (NRM) used for aperture masking interferometry (left panel) and NIRISS clear pupil mask (CLEARP) used for kernel phase imaging (right panel). The subapertures of the NRM are undersized with respect to the \emph{JWST} primary mirror segments to avoid diffraction features from the gaps between the segments. White means 100\% throughput and black means 0\% throughput.}
\label{fig:niriss_masks}
\end{figure}

The NIRISS AMI and KPI data presented here come from \emph{JWST} commissioning program 1093 (PI Deepashri Thatte). This program aimed to characterize the performance of the NIRISS non-redundant aperture mask and NIRISS kernel phase imaging with the CLEARP full pupil originally put into the NIRISS pupil wheel for engineering purposes. Data of the binary system AB~Dor as well as two point-source reference stars was taken in the AMI mode and data of three point-source reference stars was taken in the KPI mode. The throughput of the NIRISS non-redundant aperture mask is $\sim15\%$ while the throughput of the NIRISS CLEARP mask is $\sim84\%$ so that for a similarly bright target, a given number of photons is collected $\sim5.6$ times faster in the KPI mode than in the AMI mode. The optical layout of both masks is shown in Figure~\ref{fig:niriss_masks}. The observing log for the NIRISS AMI and KPI data considered here can be found in Table~\ref{tab:niriss_ami_and_kpi}.

\begin{table}[ht]
\caption{Observing log of the NIRISS AMI and KPI observations considered in this work. $T_\text{exp}$ denotes the effective integration time per exposure so that the total integration time can be obtained as $N_\text{exp}T_\text{exp}$.}
\label{tab:niriss_ami_and_kpi}
\begin{center}
\begin{tabular}{lllllll}
\rule[-1ex]{0pt}{3.5ex} Type & Observing date \& time & Target name & Filter & Mask & $N_\text{exp}$/$N_\text{int}$/$N_\text{group}$ & $T_\text{exp}$ [s] \\
\hline
\hline
\rule[-1ex]{0pt}{3.5ex} SCI & 2022-06-05T13:40:58 & AB~Dor & F480M & NRM & 2/69/5 & 26.013 \\
\rule[-1ex]{0pt}{3.5ex} REF & 2022-06-05T15:11:05 & HD~37093 & F480M & NRM & 2/61/12 & 55.193 \\
\rule[-1ex]{0pt}{3.5ex} REF & 2022-06-05T16:46:49 & HD~36805 & F480M & NRM & 2/65/9 & 44.109 \\
\hline
\rule[-1ex]{0pt}{3.5ex} SCI & 2022-06-05T19:31:40 & TYC~8906-1660-1 & F480M & CLEARP & 2/236/13 & 231.327 \\
\rule[-1ex]{0pt}{3.5ex} REF & 2022-06-05T18:50:27 & J062802.01-663738.0 & F480M & CLEARP & 2/239/14 & 252.288 \\
\rule[-1ex]{0pt}{3.5ex} REF & 2022-06-05T20:16:12 & CPD-67~607 & F480M & CLEARP & 2/232/13 & 227.406 \\
\hline
\end{tabular}
\end{center}
\end{table}

The AMI ramp (uncal) data were first processed with the \texttt{jwst} stage 1 and stage 2 pipelines. The \texttt{ipc} (inter-pixel capacitance), the \texttt{photom} (photometry), and the \texttt{resample} steps were skipped to avoid systematic errors being introduced by these steps. Instead, for NIRISS AMI and KPI observations, the target is always placed on the same detector pixel (within the target acquisition accuracy of $\sim1/3$ of a pixel) to mitigate systematic flat-fielding errors. Then, the closure phases and visibility amplitudes of AB~Dor were extracted and calibrated against those from the two point-source reference stars with \texttt{ImPlaneIA}\footnote[3]{\url{https://github.com/anand0xff/ImPlaneIA}} [\citenum{greenbaum2015}]. Finally, a binary model was fit to the calibrated data, the best fit companion was subtracted from the data, and detection limits were computed with \texttt{fouriever}\footnote[4]{\url{https://github.com/kammerje/fouriever}}. A more detailed description of the NIRISS AMI data reduction process can be found in the NIRISS AMI commissioning paper by Sivaramakrishnan et al. 2022 (in prep.).

The KPI ramp (uncal) data were first processed with the \emph{jwst} stage 1 and stage 2 pipelines in a similar fashion as the AMI data. Then, the kernel phases and their correlations were extracted with the custom \texttt{KPI3Pipeline}\footnote[5]{\url{https://github.com/kammerje/xara/tree/develop}} developed by Kammerer et al. 2022 (in prep.) which is publicly available on GitHub and currently supports NIRISS and NIRCam full pupil kernel phase observations. This pipeline is based on the \texttt{XARA}\footnote[6]{\url{https://github.com/fmartinache/xara}} package [\citenum{martinache2010},\citenum{martinache2013},\citenum{martinache2020}] and performs bad pixel corrections, windowing, and subpixel re-centering prior to extracting the kernel phases. Finally, the kernel phases of the science target were calibrated with Karhunen-Lo\`eve decomposition [\citenum{soummer2012},\citenum{kammerer2019},\citenum{wallace2020},\citenum{kammerer2021}] using the two point-source reference stars as calibrators and detection limits were computed with \texttt{fouriever}. We note that in principle, any of the three KPI targets could have been used as the science target here. A more detailed description of the NIRISS KPI data reduction process can be found in the NIRISS KPI commissioning paper by Kammerer et al. 2022 (in prep.).

\section{METHODS}
\label{sec:methods}

The successful planning, implementation, and analysis of any observing program strongly depends on the availability of user-friendly data simulation and analysis tools. Here, we aim to provide such tools for \emph{JWST} NIRCam coronagraphy to the community. The first tool is called \texttt{NIRCCoS}\footnote[7]{\url{https://github.com/kammerje/NIRCCoS}} and can be used to simulate NIRCam coronagraphy data for a given APT program based on the \texttt{pyNRC}\footnote[8]{\url{https://github.com/JarronL/pynrc}} and \texttt{WebbPSF\_ext}\footnote[9]{\url{https://github.com/JarronL/webbpsf_ext}} packages. The second tool is called \texttt{spaceKLIP}\footnote[1]{\url{https://github.com/kammerje/spaceKLIP}} and provides a \emph{JWST}-friendly wrapper for reducing and analyzing NIRCam and MIRI coronagraphy data with PSF subtraction using the KLIP [\citenum{soummer2012}] method with the \texttt{pyKLIP}\footnote[2]{\url{https://bitbucket.org/pyKLIP/pyklip/src/master/}} package [\citenum{wang2015}]. Both tools are described in the following Sections.

\subsection{Simulating NIRCam coronagraphy data with \texttt{NIRCCoS} \& \texttt{pyNRC}}
\label{sec:simulating_nircam_coronagraphy_data_with_nirccos_and_pynrc}

Simulated data plays a key role in the preparation of astronomical observations, especially for a new telescope or instrument where observers cannot rely on archival data for proposal planning. Simulated data is also vital for assessing the performance of a new telescope or instrument and validating that it is operating within the predicted range. For the new \emph{JWST}, the \texttt{MIRAGE}\footnote[3]{\url{https://github.com/spacetelescope/mirage}} data simulator can be used to simulate synthetic data from an APT file\footnote[4]{APT is the ``Astronomer's Proposal Tool'' which has to be used for proposing \emph{HST} and \emph{JWST} observations.} for a variety of NIRCam, NIRISS, and FGS observing modes. However, high-contrast imaging using coronagraphy is not supported by \texttt{MIRAGE}. To fill this gap, we developed the \texttt{NIRCCoS} tool for simulating NIRCam coronagraphy data from an APT file. The data simulation process itself is handled by \texttt{pyNRC} which was developed by Jarron Leisenring at the University of Arizona and is introduced in more detail in Girard et al. 2022 [\citenum{girard2022}]. In its most recent version, \texttt{pyNRC} can create simulated data for any NIRCam coronagraphy APT file and can consider a variety of noise sources such as photon, dark, bias, kTC, read, correlated $1/f$, IPC, post-pixel coupling, and amplifier crosstalk noise. From there, the \texttt{NIRCCoS} (``NIRCam Coronagraphy Simulations'') tool provides a user-friendly wrapper for \texttt{pyNRC} and several other \texttt{Python} tools such as \texttt{species}, \texttt{whereistheplanet}, and \texttt{jwst} to automatically gather the inputs required for injecting substellar companions into the simulated data with \texttt{pyNRC} and to run these data through the official \texttt{jwst} data reduction pipelines. The \texttt{NIRCCoS} workflow is outlined below.
\begin{enumerate}
    \item Create an APT file for your NIRCam coronagraphy program (if not already done so during the proposal planning process).
    \item Set up the YAML configuration file for your NIRCam coronagraphy program (this should take no longer than 10~minutes).
    \item If any (known) companions shall be injected into the simulated data, predict their brightness in the NIRCam bands using atmospheric models fit to literature spectrophotometry with \texttt{species} and estimate their RA and DEC offset on the date of observation based on literature astrometry using \texttt{whereistheplanet}.
    \item Simulate the ramp (uncal) data using \texttt{pyNRC}. \texttt{NIRCCoS} automatically takes care of the setup of \texttt{pyNRC} based on the provided APT and YAML configuration files and hands over the companion properties obtained in the previous step. Furthermore, background stars from the \emph{Gaia} catalog are automatically included in the simulations.
    \item If desired, run the simulated ramp data through the \texttt{jwst} stage 1, 2, and 3 coronagraphy pipelines to obtain science-ready data products (PSF-subtracted, de-rotated, and co-added images). \texttt{NIRCCoS} automatically generates the ASN association files required as an input for the \texttt{jwst} stage 3 coronagraphy pipeline.
\end{enumerate}
For each step in this workflow, there is a \texttt{Python} script that can simply be run by the user and that reads the YAML configuration file to automatically perform the described tasks.

\begin{figure}
\centering
\includegraphics[trim={2.00cm 4.75cm 2.00cm 9.25cm},clip,width=0.75\textwidth]{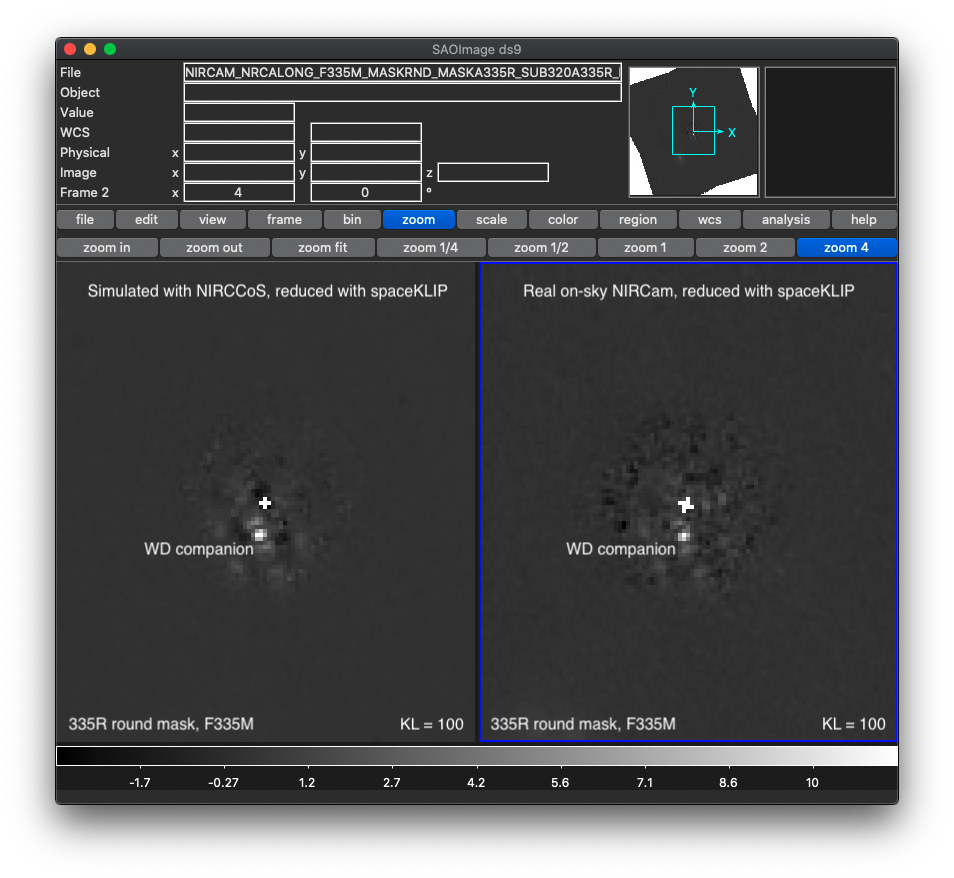}
\caption{Simulated 335R round mask F335M \texttt{NIRCCoS} image (left panel) and real on-sky NIRCam coronagraphy image (right panel) of HD~114174, both reduced with \texttt{spaceKLIP} (see Section~\ref{sec:reducing_and_analyzing_nircam_coronagraphy_data_with_spaceklip}). The host star is located in the center of the images (behind the coronagraphic mask and PSF subtracted) and the pixel scale is 63~mas. The PSF (including its wings) of the known white dwarf (WD) companion at a separation of $\sim500$~mas can be seen to the South of the host star. The color scale has been optimized to highlight the PSF of the WD companion.}
\label{fig:nirccos_vs_webb}
\end{figure}

\texttt{NIRCCoS} is publicly available including a detailed documentation with example configuration files on GitHub and has been tested extensively for simulating \emph{JWST} commissioning, GTO, and ERS observations for the high-contrast imaging ERS program [\citenum{hinkley2022}]. A qualitative comparison between the simulated \texttt{NIRCCoS} data and the real on-sky NIRCam coronagraphy data is shown in Figure~\ref{fig:nirccos_vs_webb}. A quantitative comparison can be found in Girard et al. 2022 [\citenum{girard2022}].

\subsection{Reducing \& analyzing NIRCam coronagraphy data with \texttt{spaceKLIP}}
\label{sec:reducing_and_analyzing_nircam_coronagraphy_data_with_spaceklip}

The analysis of high-contrast imaging data usually involves advanced post-processing techniques such as PSF subtraction using principal component analysis [\citenum{soummer2012},\citenum{amara2012}]. For space telescopes such as \emph{HST} and \emph{JWST}, reference star differential imaging (RDI, e.g., [\citenum{lafreniere2009},\citenum{soummer2011}]) is a common PSF subtraction method but angular differential imaging (ADI, [\citenum{marois2006}]) is also being used albeit the roll angles accessible by \emph{JWST} are restricted to a relatively small range of $\sim10$~deg on a given date for targets near the ecliptic plane due to viewing constraints.\footnote[5]{\url{https://jwst-docs.stsci.edu/jwst-observatory-characteristics/jwst-position-angles-ranges-and-offsets}} The \emph{JWST} Coronagraphic Visibility Tool\footnote[6]{\url{https://jwst-docs.stsci.edu/jwst-other-tools/jwst-target-visibility-tools/jwst-coronagraphic-visibility-tool-help}} can be used to obtain the range of position angles accessible by \emph{JWST} for a given target and on a given date and is very useful for planning coronagraphic observations with NIRCam and MIRI. The official \texttt{jwst} stage 3 coronagraphy pipeline\footnote[7]{\url{https://jwst-docs.stsci.edu/jwst-science-calibration-pipeline-overview/stages-of-jwst-data-processing/calwebb_coron3}} applies a combination of RDI and ADI and follows a custom implementation of the KLIP [\citenum{soummer2012}] method to compute the principal components of the reference star PSF library and subtract them from the science images. This implementation enables the user to change the number of subtracted KL modes (the default value is 50) but it does not enable custom choices for the number of annuli and subsections on which the Karhunen-Lo\`eve basis is evaluated. Moreover, the image registration process in the official stage 3 pipeline, which is of utmost importance for a successful principal component analysis, is very time- and memory-consuming (see below). Therefore, we decided to develop a custom stage 3 pipeline for \emph{JWST} coronagraphy data based on the popular \texttt{pyKLIP} package [\citenum{wang2015}] that offers more flexibility in selecting the PSF subtraction parameters and better performance. Furthermore, using \texttt{pyKLIP} gives access to a great variety of high-contrast imaging data analysis routines such as the computation of contrast curves and the extraction of companion photometry and astrometry using PSF forward modeling [\citenum{pueyo2016}]. The \texttt{spaceKLIP} pipeline can be used to reduce and analyze both NIRCam and MIRI coronagraphic data and is described in more detail in Carter et al. 2022 (in prep.).

\begin{figure}
\centering
\includegraphics[width=0.54\textwidth]{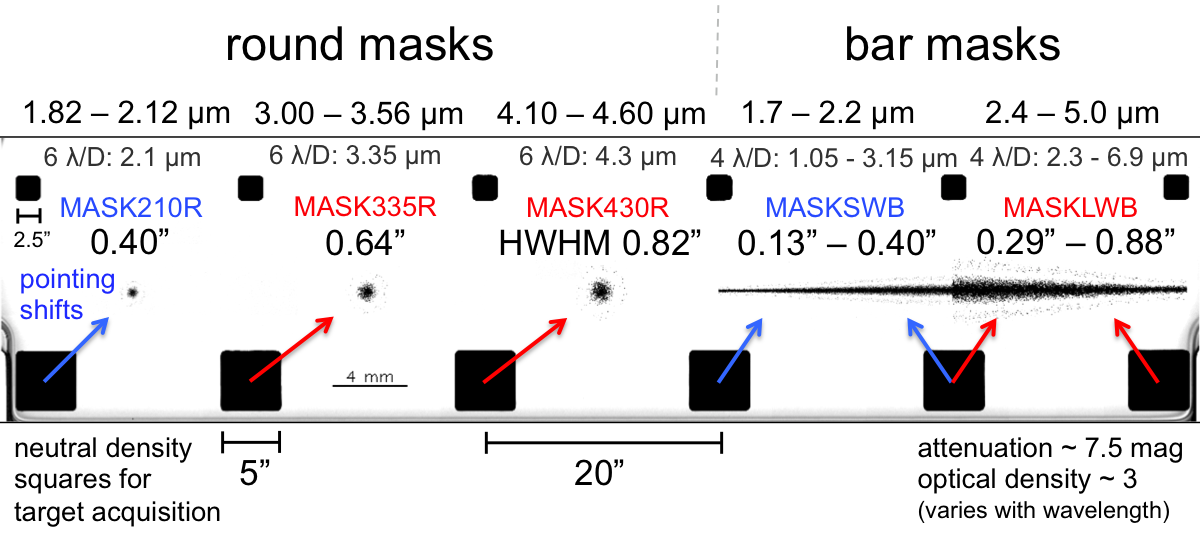}
\includegraphics[trim={0 0 0 7cm},clip,width=0.44\textwidth]{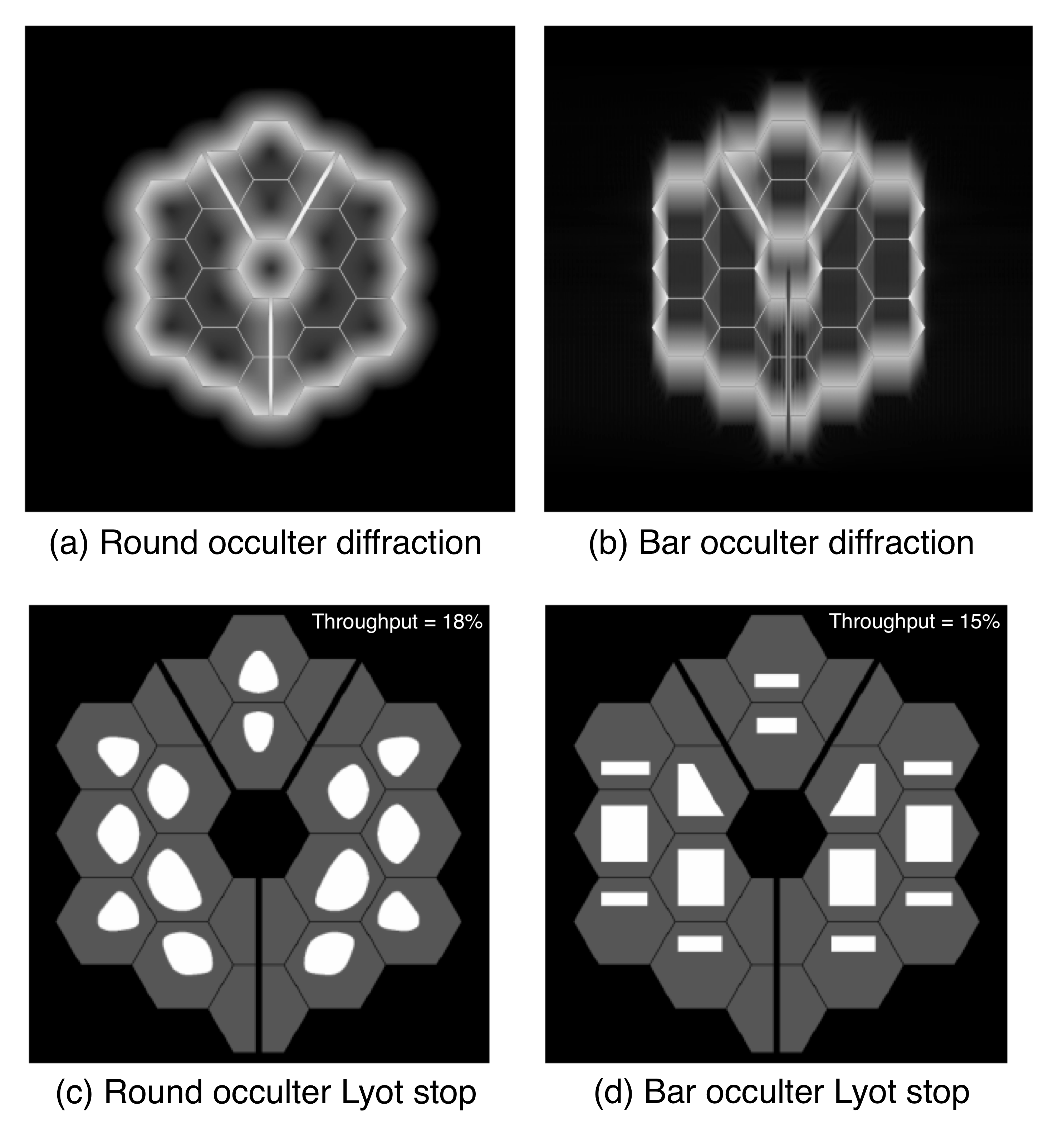}
\caption{NIRCam coronagraphic masks (left five panels) with the neutral density squares used for the target acquisition as they are located on the NIRCam coronagraphic mask holder and NIRCam Lyot stops (right two panels). In the left five panels, there are three round masks on the left and two bar masks on the right, each one optimized for a different wavelength range. In the right two panels, white means 100\% throughput and gray and black mean 0\% throughput. Taken from JDox. Adapted from Krist et al. 2010 [\citenum{krist2010}] and Mao et al. 2011 [\citenum{mao2011}].}
\label{fig:nircam_occulters}
\end{figure}

\textbf{NIRCam coronagraphic masks:} the NIRCam instrument is equipped with five different Lyot coronagraphs, three of them are round masks and two of them are bar masks [\citenum{krist2010}]. The MASK210R and the MASKSWB can only be used for short wavelength channel observations and the MASK335R, the MASK430R, and the MASKLWB can only be used for long wavelength channel observations. While the round masks are optimized for companion search and achieving the best possible contrast at large angular separations ($\gtrsim1$~arcsec), the bar masks are useful for achieving the best possible IWA and resolve companions at small angular separations. Figure~\ref{fig:nircam_occulters} has been taken from JDox\footnote[8]{\url{https://jwst-docs.stsci.edu/jwst-near-infrared-camera/nircam-instrumentation/nircam-coronagraphic-occulting-masks-and-lyot-stops}} and shows the five NIRCam coronagraphic masks with the neutral density squares used for the target acquisition as they are located on the NIRCam coronagraphic mask holder as well as the two NIRCam Lyot stops for round mask and bar mask coronagraphy. We note that the Lyot stops are quite aggressive with optical throughputs of $\sim18\%$ for the round masks and $\sim15\%$ for the bar masks to avoid light from the diffractive elements (e.g., secondary mirror support struts, gaps between mirror segments, etc.) from entering the coronagraphic instrument [\citenum{mao2011}].

\begin{figure}
\centering
\includegraphics[trim={2cm 0 2cm 0},clip,width=0.49\textwidth]{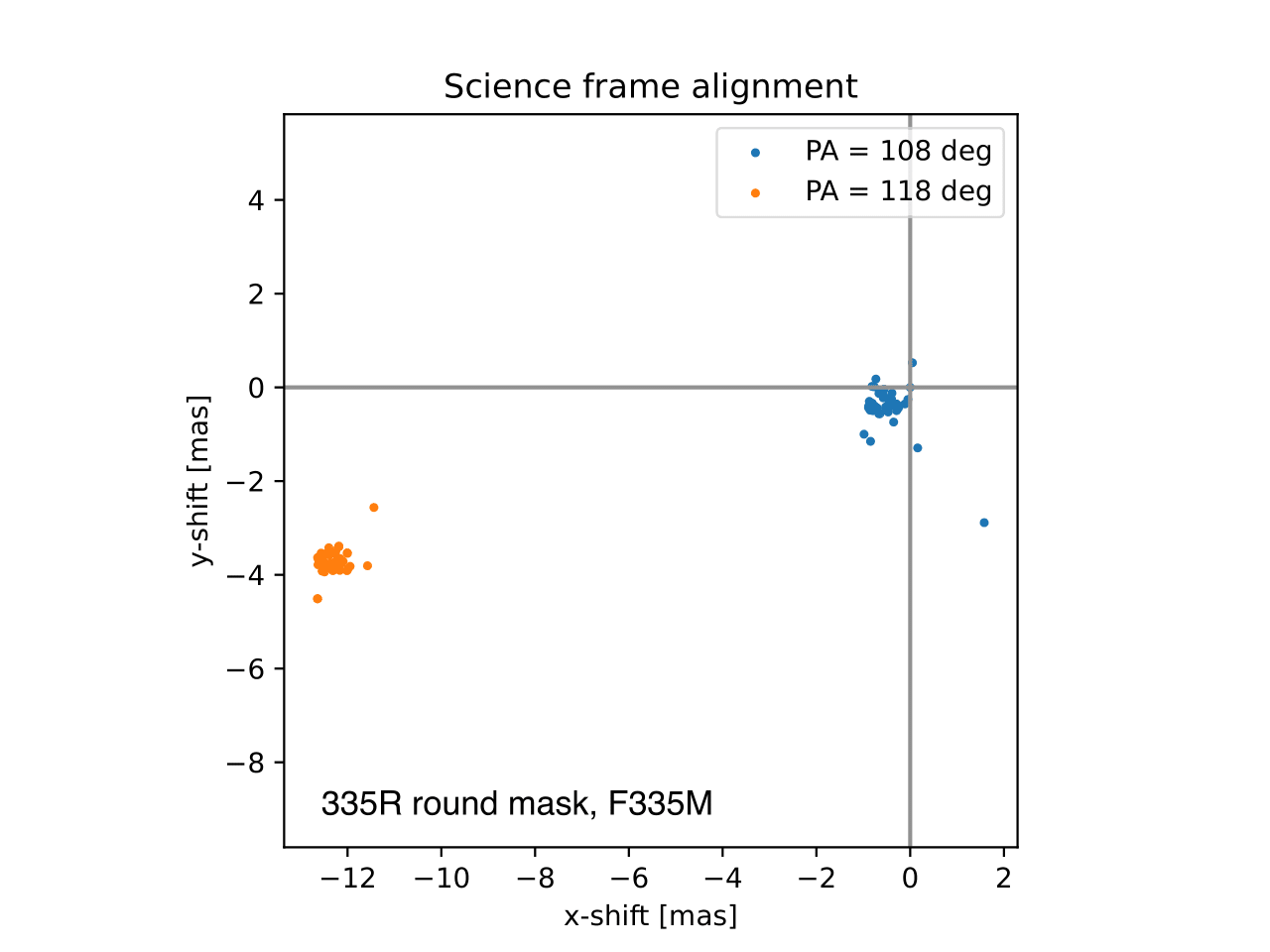}
\includegraphics[trim={2cm 0 2cm 0},clip,width=0.49\textwidth]{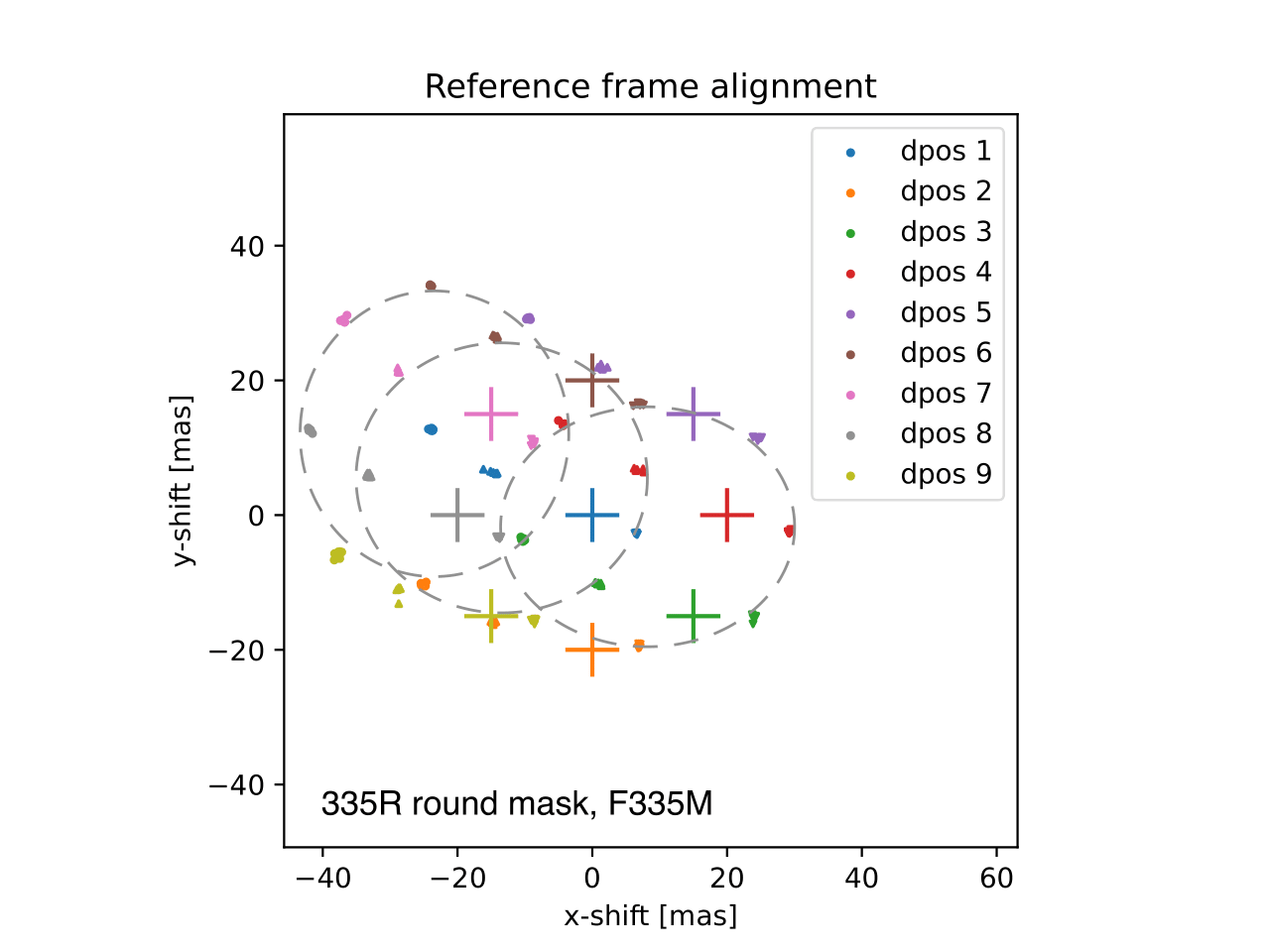}
\includegraphics[trim={2cm 0 2cm 0},clip,width=0.49\textwidth]{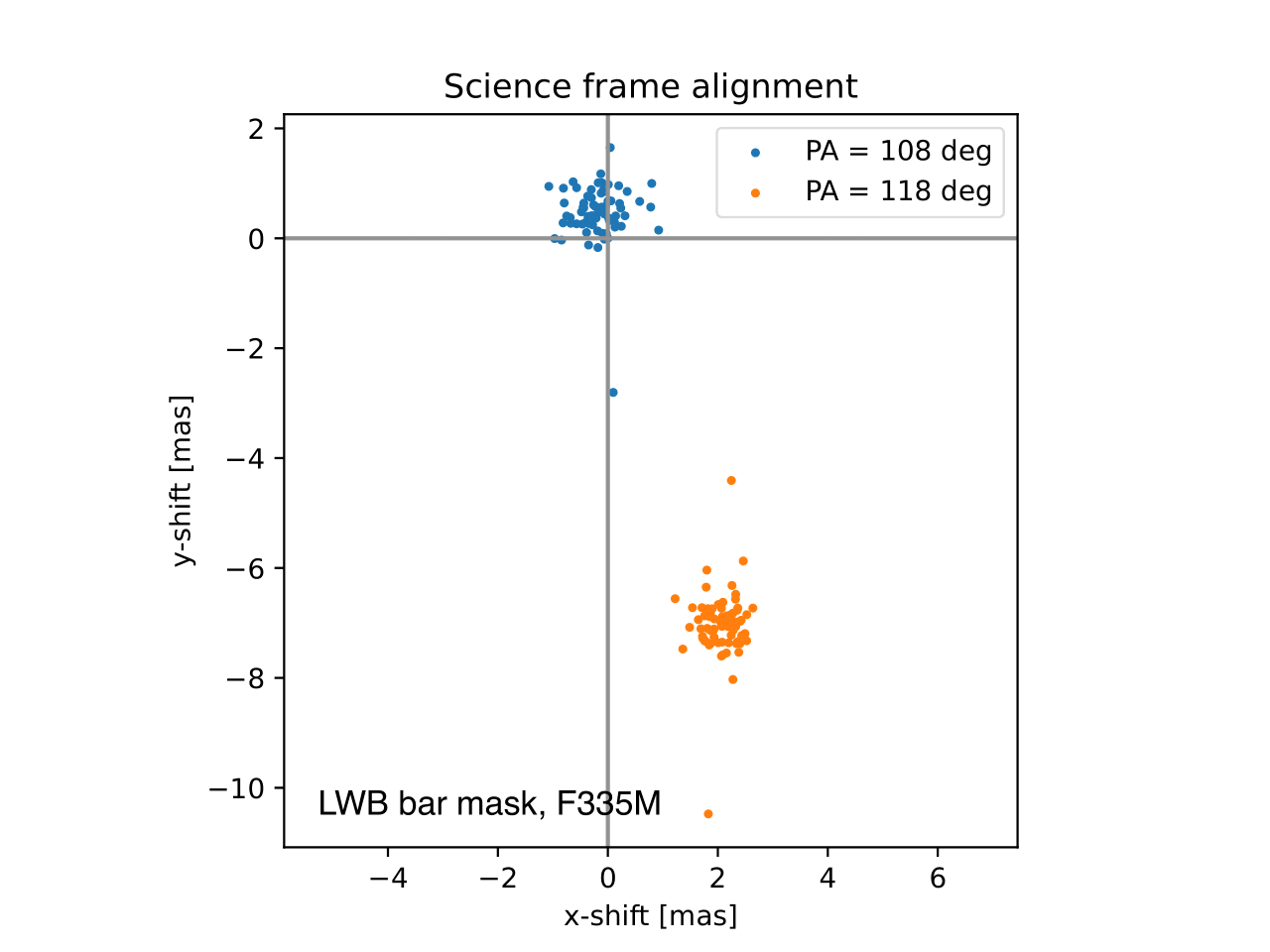}
\includegraphics[trim={2cm 0 2cm 0},clip,width=0.49\textwidth]{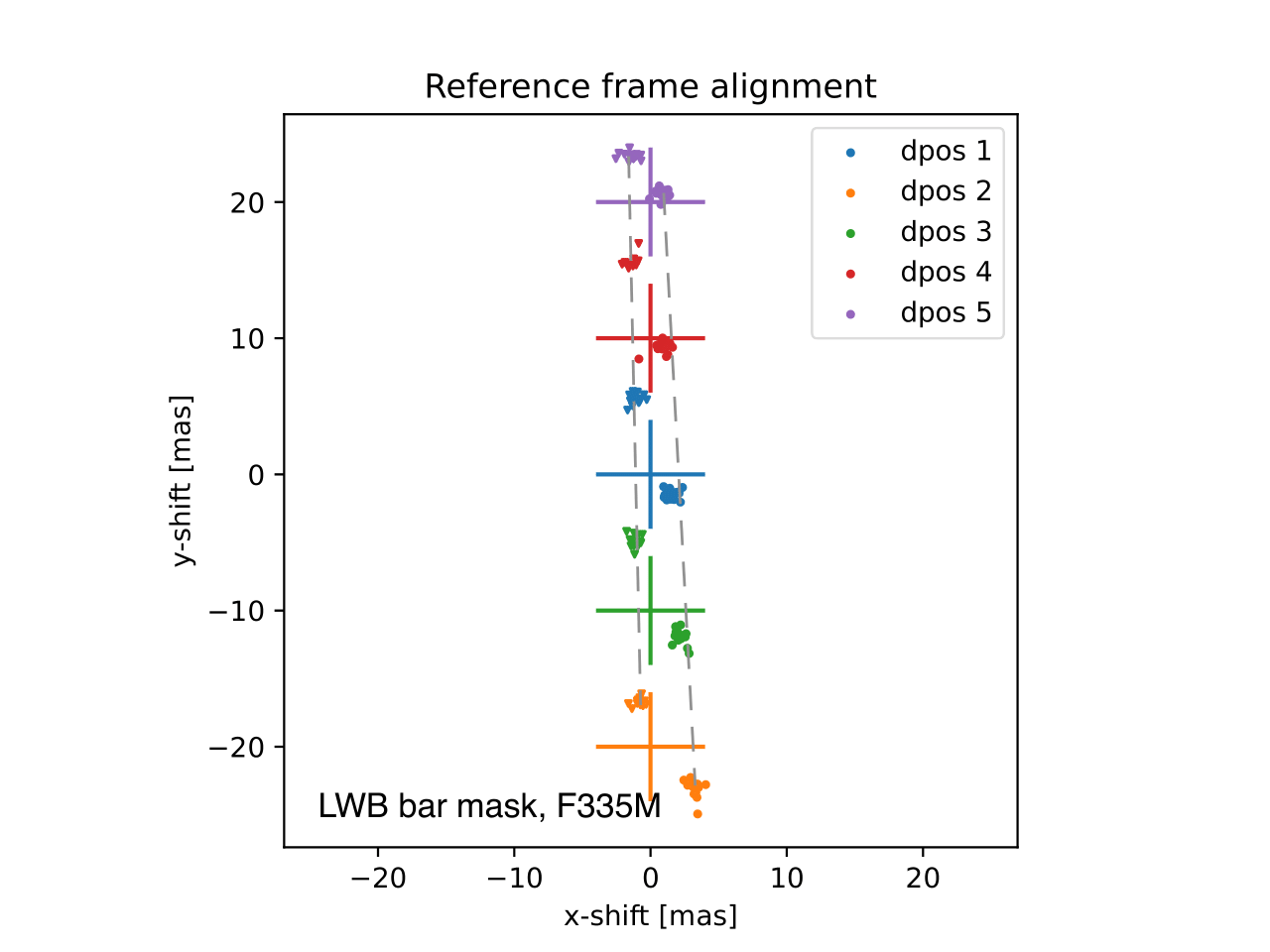}
\caption{Image registration for the NIRCam 335R round mask (top panels) and LWB bar mask (bottom panels) data of HD~114174. For each frame, the shift required to align it to the first science frame is shown. The left panels show the shifts required for the two science rolls (color-coded by the telescope roll angle) and the right panels show the shifts required for the three (round mask)/two (bar mask) reference star small-grid-dithers (SGDs, color-coded by the dither position). Frames belonging to the first, second, and third SGD are represented by circles, downward triangles, and upward triangles, respectively. In the left panels, the origin is highlighted by a gray crosshair and the first science frame falls exactly on it by construction. In the right panels, gray dashed circles and lines aid the eye in identifying the injected 9-POINT-CIRCLE (round mask) and 5-POINT-BAR (bar mask) SGDs. The color-coded crosshairs show the expected SGD positions in the absence of target acquisition errors.}
\label{fig:image_registration}
\end{figure}

\textbf{Image registration:} an important step before computing and subtracting the Karhunen-Lo\`eve principle component basis is to perform image registration and bring the science and reference PSFs to a common origin. In \texttt{spaceKLIP}, we use the same Fourier shift and subtract method that is also implemented in the official \texttt{jwst} stage 3 coronagraphy pipeline to identify the relative shift required to align two images. However, while the official pipeline registers each science frame with each reference frame resulting in a time- and memory-consuming process, \texttt{spaceKLIP} only aligns all science and reference frames to the first science frame in the provided dataset, resulting in the use of fewer computing resources. Comparisons between the products of the official pipeline and \texttt{spaceKLIP} suggest that this simplification of the image registration process does not have a significant impact on the KLIP performance. This finding is also supported by the high observatory pointing stability of generally $<1$~mas [\citenum{rigby2022}]. We note that fixing the bad pixels that were flagged by the \texttt{jwst} stage 1 pipeline prior to the image registration process has been observed to greatly improve the recovery of the injected small-grid-dithers (SGDs) and the KLIP performance, ultimately resulting in cleaner PSF-subtracted images and better detection limits at small angular separations. Due to the lack of reference pixels in the NIRCam coronagraphic subarrays, we also found that bias drifts can sometimes lead to a high fraction of flagged pixels and we reject frames for which the total number of flagged pixels exceeds the number of pixels flagged as \texttt{DO\_NOT\_USE} by a factor of more than two. We have found that this indicates a high fraction of jump-detected pixels due to the bias drifts. Figure~\ref{fig:image_registration} shows the results of the image registration process for the NIRCam round and bar mask data of HD~114174 after individual bad frames were rejected using the aforementioned method.

\textbf{Small-grid-dithers:} Soummer et al. 2014 [\citenum{soummer2014}] suggested the use of SGDs with strokes of a few tens of milliarcseconds to build up a reference PSF library with sufficient spatial diversity to sample typical \emph{JWST} small angle maneuver target acquisition errors. Following their suggestion, a number of SGDs were implemented for NIRCam coronagraphy observations\footnote[9]{\url{https://jwst-docs.stsci.edu/jwst-near-infrared-camera/nircam-operations/nircam-dithers-and-mosaics/nircam-subpixel-dithers/nircam-small-grid-dithers}}. Considering the relative offsets observed between two science rolls or between different reference star observations in Figure~\ref{fig:image_registration}, we find an in-flight small angle maneuver target acquisition error of a few tens of milliarcseconds. This finding is also supported by telescope pointing telemetry collected by the Fine Guidance Sensor. Girard et al. 2022 [\citenum{girard2022}] have shown that using a reference star observation separated by $\sim40$~mas from the science target observation results in a degradation of the contrast performance of $\sim1$ order of magnitude (i.e., a factor of $\sim10$) over using a reference star observation separated by only $\sim10$~mas. For the time being, we therefore highly recommend the use of SGDs for NIRCam coronagraphy reference star observations. However, we note that commissioning performance is preliminary and there is ongoing work to continue improving the astrometric calibrations and the target acquisition procedure for NIRCam.

\begin{figure}
\centering
\includegraphics[width=0.50\textwidth]{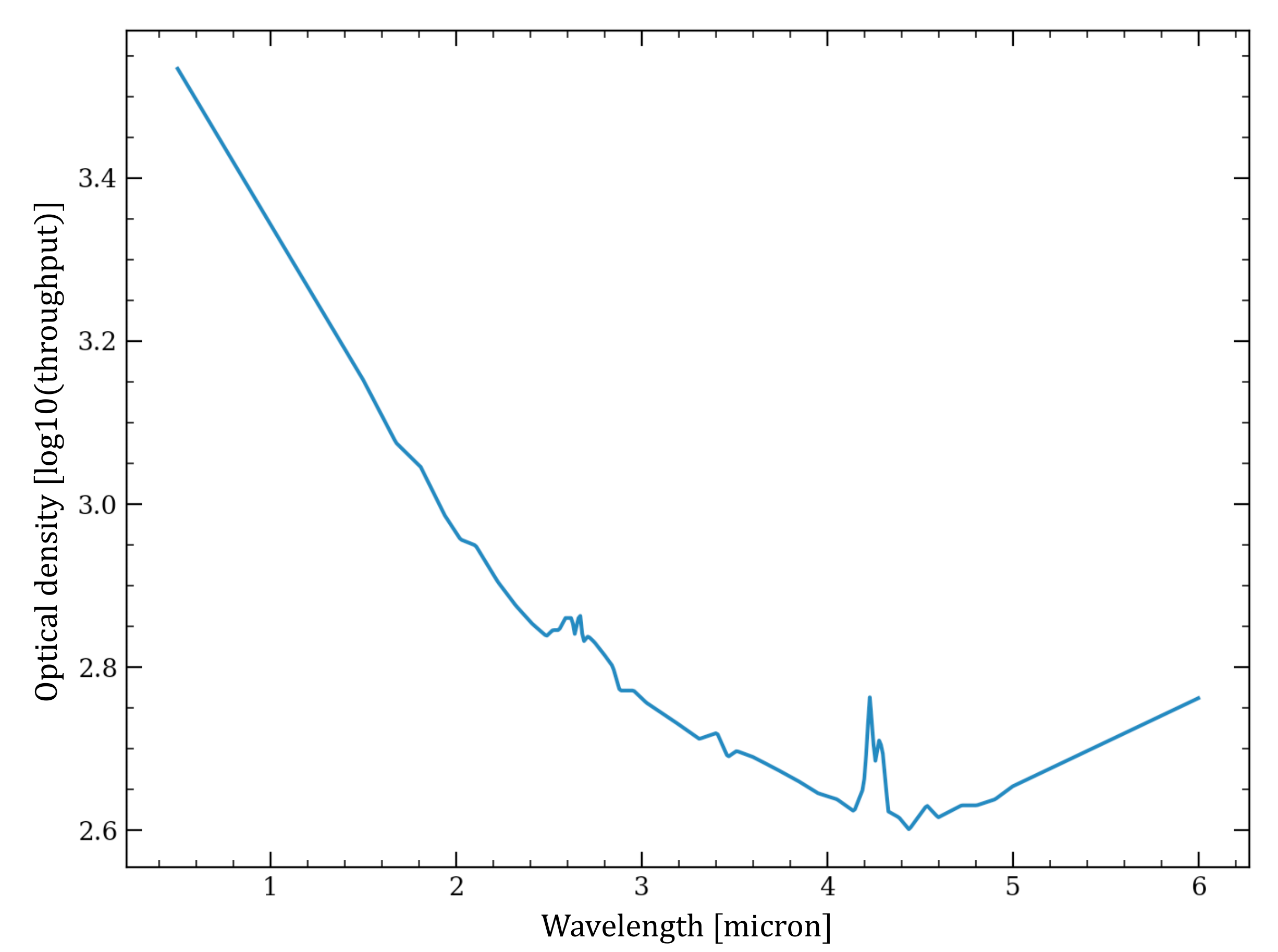}
\caption{Chromaticity of the optical density of the NIRCam neutral density squares used for the target acquisition of bright targets (K $<$ 6.3~mag for short wavelength channel observations and K $<$ 4.7~mag for long wavelength channel observations). The data was measured during the Cryo-Vac 3 (CV3) Thermal Vacuum Test [\citenum{kimble2016}] on the ground.}
\label{fig:optical_density}
\end{figure}

\textbf{Contrast curves:} after performing the RDI, ADI, or RDI+ADI PSF subtraction using the KLIP [\citenum{soummer2012}] method, \texttt{spaceKLIP} can be used to compute contrast curves from the PSF-subtracted images. Since the brightness of the host star can hardly be measured in the coronagraphic images due to the presence of the focal plane mask and since \emph{JWST} has no satellite spots for the photometric calibration such as VLT/SPHERE for instance [\citenum{beuzit2019}], we use the unocculted target acquisition image of the science target, observed through one of the neutral density squares if the target is bright or through a clear aperture if it is faint, for correctly normalizing the coronagraphic contrast curves. We note that the optical density of the neutral density squares is strongly wavelength-dependent (see Figure~\ref{fig:optical_density}) and we here use the average of the optical density over the filter bandpass. In a future version of \texttt{spaceKLIP}, the optical density should be weighted by the filter curve and the host star spectrum to achieve a higher photometric precision. Target acquisition images are taken in the F210M filter for short wavelength channel observations and in the F335M filter for long wavelength channel observations\footnote[1]{\url{https://jwst-docs.stsci.edu/jwst-near-infrared-camera/nircam-operations/nircam-target-acquisition/nircam-coronagraphic-imaging-target-acquisition}}. In case the science images were taken in another filter, we fit a PHOENIX stellar model [\citenum{husser2013}] to the public VizieR\footnote[2]{\url{http://vizier.u-strasbg.fr/vizier/sed/}} photometry of the host star and scale the brightness measured in the target acquisition image to the respective filter using this stellar model. Finally, we normalize the contrast curves by the coronagraphic mask throughput.

\begin{figure}
\centering
\includegraphics[trim={17cm 0 2cm 0},clip,width=0.49\textwidth]{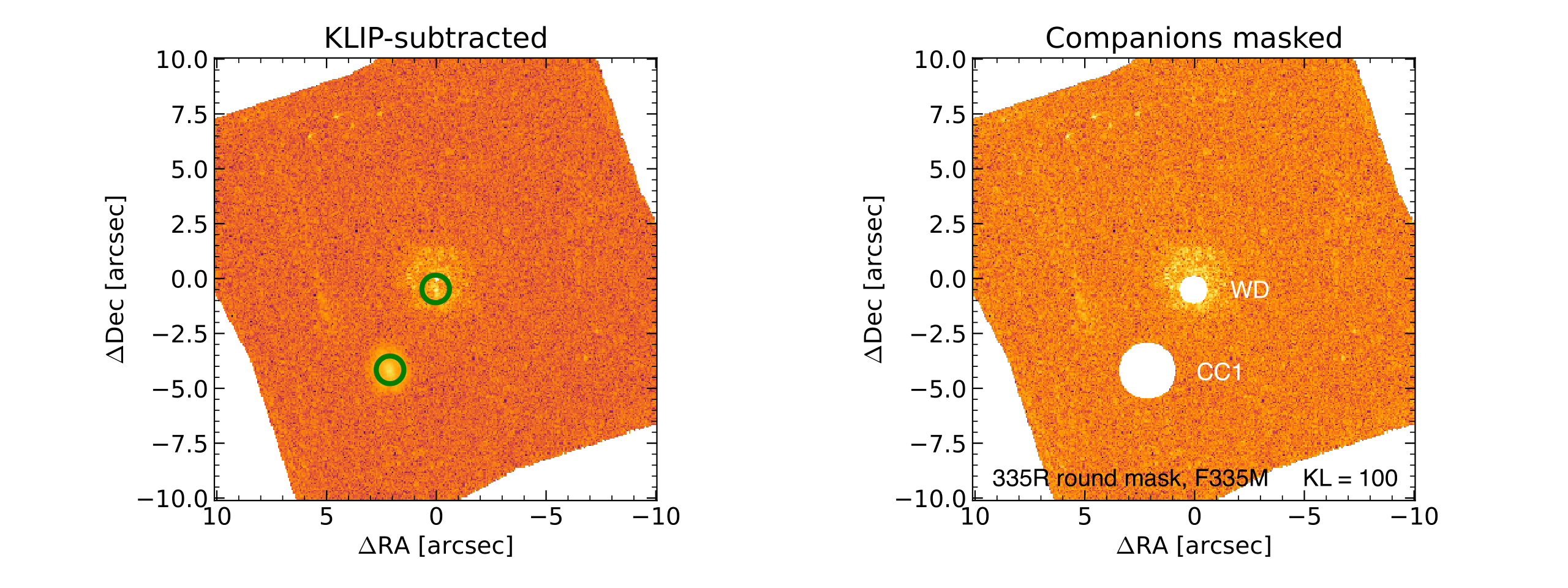}
\includegraphics[trim={17cm 0 2cm 0},clip,width=0.49\textwidth]{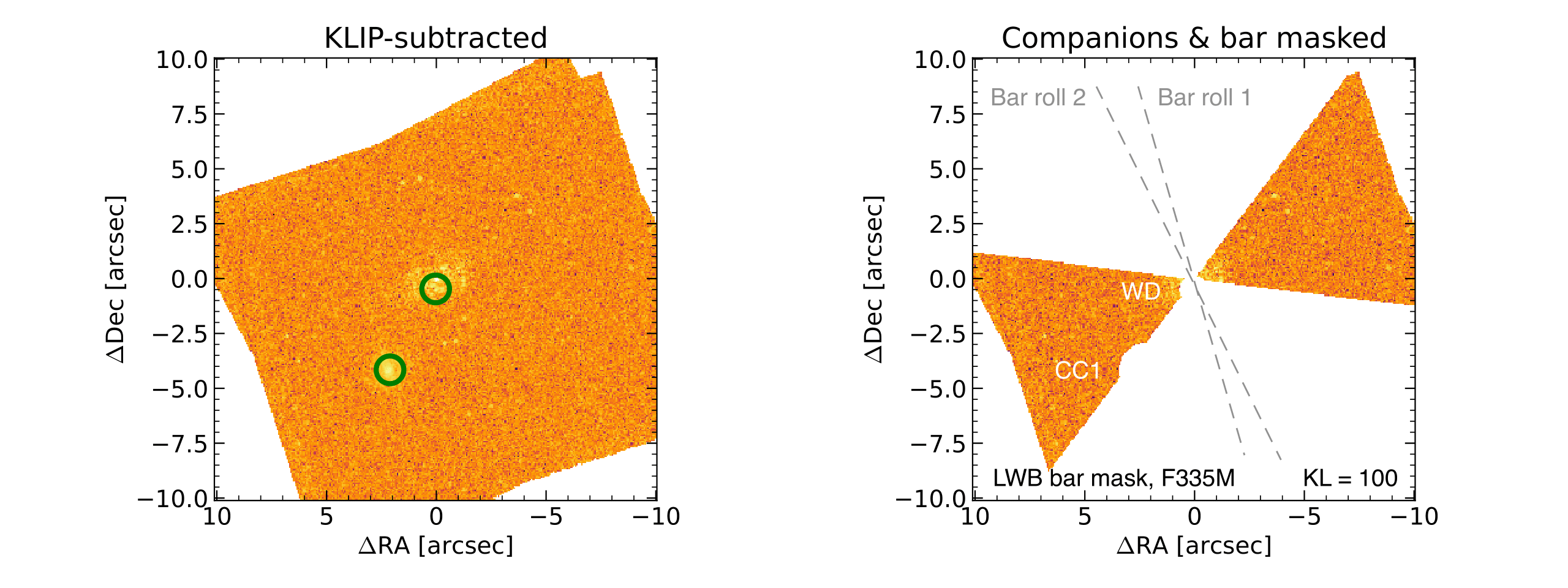}
\caption{Final KLIP-subtracted F335M 335R round mask (left panel) and LWB bar mask (right panel) images of HD~114174. In both images, the entire PSF including the PSF wings of a companion candidate (CC1, likely a background source) has been masked out with a $r = 12~\lambda/D$ mask and the PSF core of the known white dwarf (WD) companion has been masked out with a smaller $r = 6~\lambda/D$ mask because its PSF wings are very faint and we do not want to degrade the inner working angle. In the bar mask image (right panel), the approximate location of the bar in each of the two rolls is highlighted by gray dashed lines and the contrast curve is only computed from the regions perpendicular to the bar. We here only consider a narrow region around the axis perpendicular to the bar in order to achieve the best possible IWA. In both images, the number of subtracted KL modes is 100 and the residual speckle halo of the host star can be seen as a bright feature in the center.}
\label{fig:companion_and_bar_masking}
\end{figure}

\textbf{Companion \& bar masking:} when computing contrast curves, it is often desirable to mask out any detected companions to avoid them from biasing the contrast measurement. This can be achieved in \texttt{spaceKLIP} by specifying the approximate positions of these companions in the configuration file. While the coronagraphic throughput of the round masks is azimuthally symmetric around the mask center, a further subtlety with the bar masks is that the coronagraphic throughput perpendicular to the bar is very different to that along the bar (which is basically zero, see Figure~\ref{fig:nircam_occulters}). Hence, the region occulted by the bar in either of the rolls should be masked out and ignored when computing contrast curves. This can be achieved in \texttt{spaceKLIP} by specifying the ranges of position angles that shall be considered when computing contrast curves. Figure~\ref{fig:companion_and_bar_masking} shows the final KLIP-subtracted 335R round mask and LWB bar mask images of HD~114174 with the detected companions and the regions occulted by the bar in either of the rolls masked out. We here only consider a narrow region around the axis perpendicular to the bar in order to achieve the best possible IWA.

\begin{figure}
\centering
\includegraphics[trim={0 0 0 1cm},clip,width=\textwidth]{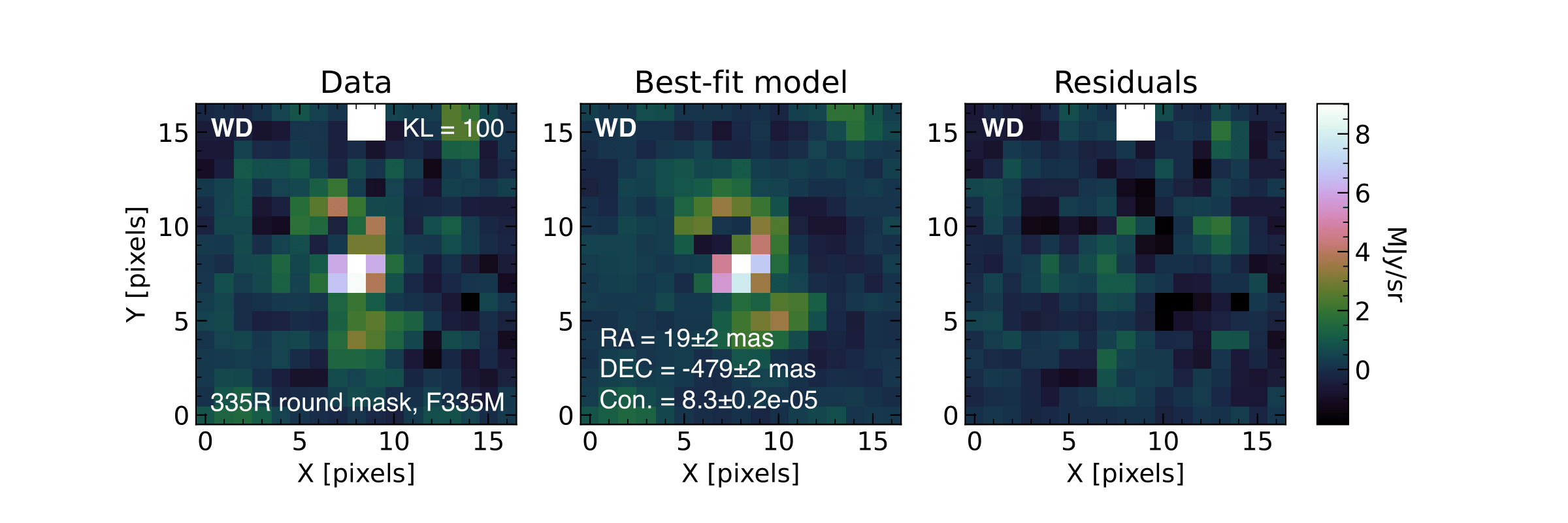}
\includegraphics[trim={0 0 0 1cm},clip,width=\textwidth]{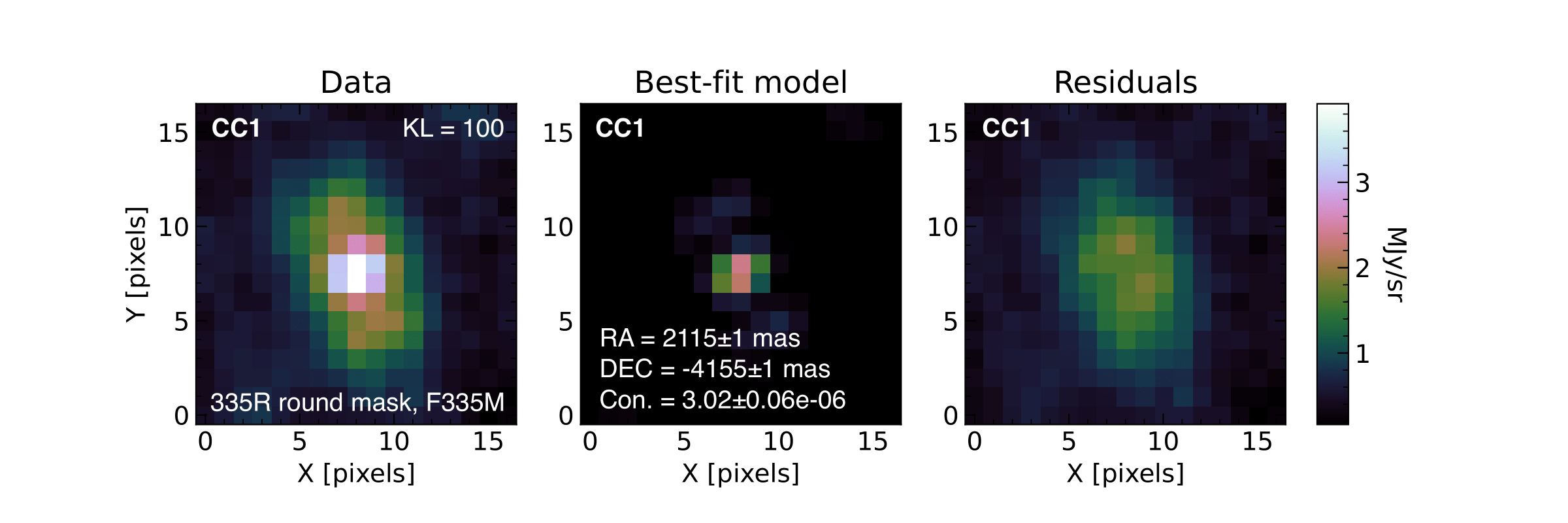}
\caption{KLIP-subtracted 335R round mask PSF (left panels), forward-modeled PSF from \texttt{spaceKLIP} (middle panels), and residuals (right panels) for the known WD companion (top panels) and the CC1 companion candidate (bottom panels) in the HD~114174 data. While the forward-modeled PSF of the WD companion gives a good fit to the data, the one of the CC1 companion candidate leaves a significant resolved halo in the residuals. Given that the fit with a point-source PSF is unsatisfactory, we suspect that the CC1 companion candidate is a resolved background source (e.g., a faint galaxy).}
\label{fig:companion_forward_modeling}
\end{figure}

\textbf{Companion forward modeling:} the photometry and astrometry of any detected companions can be extracted with \texttt{spaceKLIP} using PSF forward modeling. We simulate an unocculted off-axis PSF using \texttt{WebbPSF} and apply the position-dependent coronagraphic mask throughput during the forward modeling process of the companion PSF. The unocculted off-axis PSF is normalized in the same fashion as the contrast curves. We then obtain the best fit contrast and relative astrometry of the companion using a Monte-Carlo Markov Chain minimization routine on the residuals between the final KLIP-subtracted images and the forward-modeled companion PSF. Figure~\ref{fig:companion_forward_modeling} shows the forward modeling results for the known WD companion and the CC1 companion candidate in the HD~114174 data. In a future version of \texttt{spaceKLIP}, a position-dependent off-axis PSF evaluated at the location of the companion should be obtained using \texttt{WebbPSF\_ext} and used instead of the completely unocculted off-axis PSF from \texttt{WebbPSF} for higher fidelity.

\section{RESULTS \& DISCUSSION}
\label{sec:results_and_discussion}

\begin{figure}
\centering
\includegraphics[width=\textwidth]{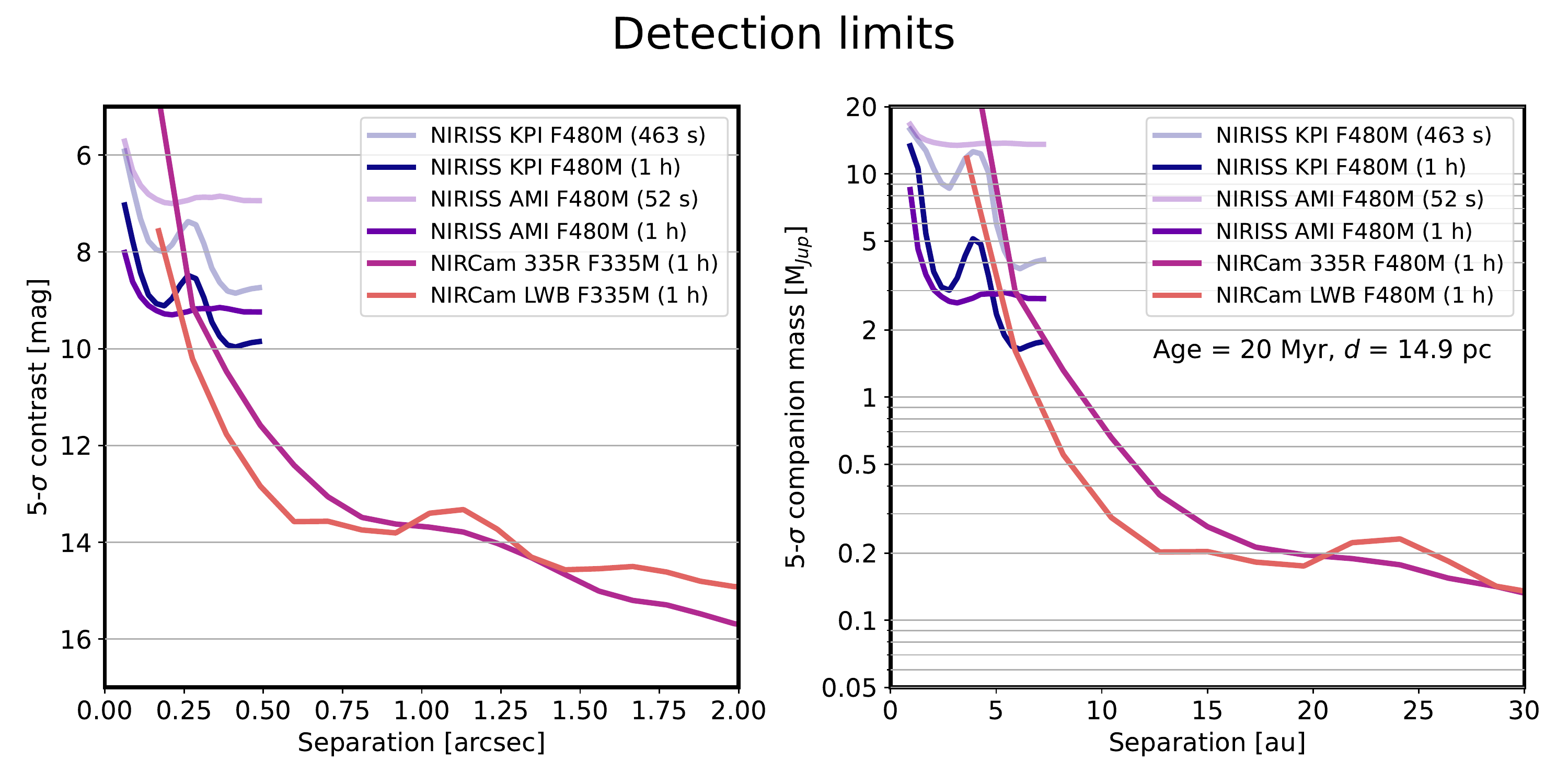}
\caption{Contrast limits as a function of the angular separation from the host star for the NIRCam 335R round and LWB bar coronagraphic masks and the NIRISS AMI and KPI observing modes (left panel). All contrast curves are scaled to a total integration time of 1~hour and a target of the brightness of AB~Dor assuming that all observations are photon noise-limited. Companion mass limits for an AB~Dor-like star at an age of 20~Myr and a distance of $\sim14.9$~pc (right panel). The NIRCam coronagraphy F335M contrast curves were translated to the F480M filter assuming that they are limited by speckle noise and that the angular resolution scales linearly with wavelength.}
\label{fig:detection_limits}
\end{figure}

Companion detection limits as a function of the angular separation from the host star are a widespread metric for the performance of a high-contrast imaging instrument or mode. These limits depend strongly on the host star brightness and the total exposure time, or more directly on the number of photons collected on the detector. Figure~\ref{fig:detection_limits} compares the companion detection limits for the NIRCam 335R round and LWB bar coronagraphic masks with the NIRISS AMI and KPI observing modes. For a more direct comparison, all contrast curves are scaled to a total integration time of 1~hour and a target of the brightness of AB~Dor assuming that all observations are photon noise-limited. This assumption should be reasonable as long as the detection limits are above the systematic noise floor estimated to be around $\sim10$~mag contrast for NIRISS AMI and KPI [\citenum{greenbaum2015}] and $\sim16$~mag contrast for NIRCam coronagraphy [\citenum{perrin2018}]. The detection limits for the NIRISS AMI and KPI observing modes are only shown out to 500~mas separation since the field-of-view of these techniques is limited to $\lambda/B_\text{min} \approx 260$--330~mas (depending on the filter wavelength $\lambda$ and the minimum baseline $B_\text{min}$) in the sense that aliasing and model redundancies will become an issue at larger separations.

\begin{table}[ht]
\caption{\emph{WISE} [\citenum{wright2010}] W1-band ($\sim3.4~\text{\textmu m}$) and W2-band ($\sim4.6~\text{\textmu m}$) photometry of the targets considered in this work.}
\label{tab:contrast_curve_scaling}
\begin{center}
\begin{tabular}{lllllll}
\rule[-1ex]{0pt}{3.5ex} Observing mode & Filter & Target & $T_\text{tot}$ [s] & W1 [mag] & W2 [mag] \\
\hline
\hline
\rule[-1ex]{0pt}{3.5ex} NIRISS KPI & F480M & TYC~8906-1660-1 & 463 & 8.239 & 8.340 \\
\rule[-1ex]{0pt}{3.5ex} NIRISS AMI & F480M & AB~Dor & 52 & 4.419 & 3.885 \\
\rule[-1ex]{0pt}{3.5ex} NIRCam 335R & F335M & HD~114174 & 5134 & 5.214 & 5.066 \\
\rule[-1ex]{0pt}{3.5ex} NIRCam LWB & F335M & HD~114174 & 3057 & 5.214 & 5.066 \\
\hline
\end{tabular}
\end{center}
\end{table}

Figure~\ref{fig:detection_limits} shows the 5--$\sigma$ companion detection limits for NIRCam 335R round and LWB bar mask coronagraphy, NIRISS AMI, and NIRISS KPI. All contrast curves are scaled to a total integration time of 1~hour and a target of the brightness of AB~Dor assuming that all observations are photon noise-limited and scale with $\sqrt{N_\text{phot}}$, where $N_\text{phot}$ is the number of detected photons. This means that the estimated detection limits in 1~hour are given as the product of the measured detection limits and a scaling factor. This scaling factor is computed in each of the different cases as follows:
\begin{enumerate}
    \item the shown NIRISS AMI 52~s detection limits are the ones measured from the data presented in Section~\ref{sec:observations} and the 1~hour detection limits are obtained by scaling the 52~s ones with \begin{equation*}
        \sqrt{52/3600} \approx 0.120,
    \end{equation*}
    \item the shown NIRISS KPI 463~s detection limits are the ones measured from the data presented in Section~\ref{sec:observations} but scaled by
    \begin{equation*}
        \sqrt{10^{(3.885-8.340)/2.5}} \approx 0.129
    \end{equation*}
    which is the brightness difference between AB~Dor and TYC~8906-1660-1 at $\sim$F480M and the 1~hour detection limits are obtained by further scaling the shown 463~s ones with
    \begin{equation*}
        \sqrt{463/3600} \approx 0.359,
    \end{equation*}
    \item the shown NIRCam 335R round mask 1~hour detection limits are the ones measured from the data presented in Section~\ref{sec:observations} but scaled by
    \begin{equation*}
        \sqrt{10^{(4.419-5.214)/2.5}}\sqrt{5134/3600} \approx 0.828
    \end{equation*}
    to account for the brightness difference between AB~Dor and HD~114174 at $\sim$F335M and the integration time difference to 1~hour,
    \item the shown NIRCam LWB bar mask 1~hour detection limits are the ones measured from the data presented in Section~\ref{sec:observations} but scaled by
    \begin{equation*}
        \sqrt{10^{(4.419-5.214)/2.5}}\sqrt{3057/3600} \approx 0.639
    \end{equation*}
    to account for the brightness difference between AB~Dor and HD~114174 at $\sim$F335M and the integration time difference to 1~hour.
\end{enumerate}
The NIRCam 335R round and LWB bar masks perform very similar at separations between $\sim0.75$--1.5~arcsec. Further out, the 335R round mask delivers better performance and it should be noted that the round mask gives access to the full 360~deg field-of-view while the bar mask suffers from low throughput in the regions along the bar, significantly restricting the search space for new companions. Inside of $\sim750$~mas, the bar mask performs slightly better due to its reduced IWA. Even further inside of $\sim250$~mas, Fourier plane imaging techniques such as AMI and KPI are required which achieve contrasts of $\sim9$~mag at separations of $\sim200$~mas and $\sim7$--8~mag at separations of $\sim70$~mas. In the regime where Fourier techniques are superior to coronagraphy (i.e., $\lesssim250$~mas), AMI slightly outperforms KPI for a very bright target such as AB~Dor (cf., [\citenum{sallum2019}]). However, it should be noted that KPI can still be used for faint targets that cannot be observed with the non-redundant mask due to its much lower throughput if compared to full pupil kernel phase imaging.

%For comparison with detection limits achieved from the ground, Figure~\ref{fig:detection_limits} also shows a Keck NIRC2 L-band contrast curve from Thompson et al. 2022 (submitted). The data from which this contrast curve was computed were acquired without a coronagraph. Thompson et al. 2022 (submitted) found non-coronagraphic sequences outperforming Lyot coronagraphy below a few $\lambda/D$ separation. Due to its larger primary mirror ($\sim10$~m for Keck instead of $\sim6.5$~m for \emph{JWST}), NIRC2 achieves a smaller IWA and a similar contrast performance between $\sim0.2$--1.25~arcsec if compared to NIRCam coronagraphy albeit having to cope with thermal background noise from the Earth's atmosphere. Nevertheless, if considering the $>2$ times smaller collecting area of \emph{JWST} if compared to Keck and the contrast performance at large separations of $>1.25$~arcsec, the NIRCam coronagraphy performance measured during the instrument commissioning clearly surpasses the best possible limits achieved with ground-based L-band imaging.

For evaluating the scientific capabilities of \emph{JWST} high-contrast imaging, we converted the observed 5--$\sigma$ contrast curves into absolute magnitude curves by generating synthetic photometry of a K0V BT-NextGen model [\citenum{allard2012}] fit to the 2MASS photometry of AB~Dor at a distance of $\sim14.9$~pc [\citenum{gaia2022}] using the \texttt{species} package [\citenum{stolker2020}]. We then converted the absolute magnitude curves into companion mass limits for each instrument using a combination of \texttt{ATMO} (for high masses, [\citenum{phillips2020}]) and \texttt{BEX} (for low masses, [\citenum{linder2019}]) evolutionary models at an age of 20~Myr and a distance of $\sim14.9$~pc based on the methodology described in Carter et al. 2021 [\citenum{carter2021}]. The interpolated models were averaged where they overlap, and the \texttt{BEX} models were re-computed over the \emph{JWST} bandpasses considered here. We then converted angular separation in arcsec into orbital separation in au assuming the \textit{Gaia} DR3 parallax measured for AB~Dor [\citenum{gaia2022}]. Figure \ref{fig:detection_limits} shows the computed companion mass limits for each instrument and observing mode considered here. The NIRCam coronagraphy F335M contrast curves were translated to the F480M filter assuming that they are limited by speckle noise and that the angular resolution scales linearly with wavelength. We did this because the F335M commissioning filter is not the best \emph{JWST} filter to detect the lowest mass companions due to a deep methane absorption feature. We note that the NIRCam coronagraphy companion mass limits might hence overpredict the performance at large separations. We find that NIRISS AMI and KPI at $4.8~\text{\textmu m}$ are sensitive to $\sim2$--3 Jupiter mass companions at small separations, and NIRCam coronography is sensitive to $\sim0.2$ Jupiter mass companions between 10--25~au.

\section{SUMMARY \& CONCLUSIONS}
\label{sec:summary_and_conclusions}

In this paper, we present our publicly available \texttt{Python} tools \texttt{NIRCCoS}\footnote[3]{\url{https://github.com/kammerje/NIRCCoS}} and \texttt{spaceKLIP}\footnote[4]{\url{https://github.com/kammerje/spaceKLIP}} that can be used to simulate, reduce, and analyze \emph{JWST} NIRCam coronagraphy data. These tools will be helpful for Cycle 1 observers to obtain science-ready data products such as companion detection limits and detected companion properties from their own programs as well as for Cycle 2 proposal planning. We note that \texttt{spaceKLIP} can also be used to reduce and analyze MIRI coronagraphy data (see Carter et al. 2022 in prep.).

We furthermore present companion detection limits computed from on-sky \emph{JWST} commissioning observations with the NIRCam 335R round and LWB bar coronagraphic masks and compare them to NIRISS AMI and KPI detection limits also measured during the instrument commissioning (see NIRISS AMI and KPI commissioning papers by Sivaramakrishnan et al. 2022 in prep. and Kammerer et al. 2022 in prep., respectively). We note that for the NIRCam coronagraphy data, especially the bar mask data, a robust bad pixel cleaning and image registration procedure is essential to achieve the detection limits presented here. Within 1~hour of integration time and for a bright target such as AB~Dor, we find that NIRCam coronagraphy reaches contrasts of $\sim13$~mag at $\sim500$~mas and $\sim15$~mag at $\sim2$~arcsec. Under similar conditions, NIRISS AMI and KPI reach contrasts of $\sim7$--8~mag at $\sim70$~mas and $\sim9$~mag at $\sim200$~mas. Based on our analysis of NIRCam and NIRISS commissioning data, we recommend to
\begin{enumerate}
    \item use NIRISS AMI for high-contrast imaging inside of $\sim250$~mas around bright targets,
    \item use NIRISS KPI for high-contrast imaging inside of $\sim250$~mas around faint targets,
    \item use NIRCam coronagraphy for high-contrast imaging outside of $\sim250$~mas.
\end{enumerate}

Finally, we remind the reader that this work presents an early look at the initial performance of \emph{JWST} NIRCam coronagraphy and NIRISS AMI and KPI. The achieved performance may naturally evolve as experience is gained. We also note that even though the instrument commissioning is formally completed, instrument calibration activities will continue throughout Cycle 1.

\acknowledgments % equivalent to \section*{ACKNOWLEDGMENTS}       
 
These observations were made possible through the efforts of the many hundreds of people in the international commissioning team for JWST. This work is based on observations made with the NASA/ESA/CSA James Webb Space Telescope. The data were obtained from the Mikulski Archive for Space Telescopes at the Space Telescope Science Institute, which is operated by the Association of Universities for Research in Astronomy, Inc., under NASA contract NAS 5-03127 for JWST. These observations are associated with program \#1441 and \#1093. Support for programs \#1194, \#1411, and \#1412 was provided by NASA through a grant from the Space Telescope Science Institute, which is operated by the Association of Universities for Research in Astronomy, Inc., under NASA contract NAS 5-03127.

% References
\bibliography{report} % bibliography data in report.bib
\bibliographystyle{spiebib} % makes bibtex use spiebib.bst

\end{document}